\documentclass[%
 reprint,
superscriptaddress,
 amsmath,amssymb,
 aps,
]{revtex4-2}
\usepackage[ruled,vlined,linesnumbered]{algorithm2e}
\usepackage[colorlinks=true, allcolors=blue]{hyperref}
\usepackage{graphicx}% Include figure files
\usepackage{dcolumn}% Align table columns on decimal point
\usepackage{bm}% bold math
\usepackage{float}

\begin{document}
\raggedbottom
% \preprint{APS/123-QED}

\title{Quantum Statistics from Classical Simulations via Generative Gibbs Sampling}

\author{Weizhou Wang}
\affiliation{Department of Chemistry, University of Chicago, Chicago, Illinois 60637, USA}
\affiliation{James Franck Institute, University of Chicago, Chicago, Illinois 60637, USA}

\author{Xuanxi Zhang}
\affiliation{Courant Institute of Mathematical Sciences, New York University, New York, New York 10012, USA}

\author{Jonathan Weare}
\email{weare@nyu.edu}
\affiliation{Courant Institute of Mathematical Sciences, New York University, New York, New York 10012, USA}

\author{Aaron R. Dinner}
\email{dinner@uchicago.edu}
\affiliation{Department of Chemistry, University of Chicago, Chicago, Illinois 60637, USA}
\affiliation{James Franck Institute, University of Chicago, Chicago, Illinois 60637, USA}

\date{\today}

\begin{abstract}

Accurate simulation of nuclear quantum effects is essential for molecular modeling but expensive using path integral molecular dynamics (PIMD). We present GG-PI, a ring-polymer-based framework that combines generative modeling of the single-bead conditional density with Gibbs sampling to recover quantum statistics from classical simulation data. GG-PI uses inexpensive standard classical simulations or existing data for training and allows transfer across temperatures without retraining. On standard test systems, GG-PI significantly reduces wall clock time compared to PIMD. Our approach extends easily to a wide range of problems with similar Markov structure.
\end{abstract}

\maketitle

\section{Introduction}
Molecular dynamics (MD) is indispensable in physics, chemistry, and biology. However, in its standard form, its classical treatment of nuclei neglects nuclear quantum effects (NQEs) \cite{markland_nuclear_2018}. By factorizing the imaginary-time propagator, path-integral molecular dynamics (PIMD) \cite{ceperley_path_1995,tuckerman_efficient_1993} and path-integral Monte Carlo (PIMC) \cite{metropolis1953,hastings1970,herman_path_1982} establish an isomorphism between quantum statistics and a classical ring polymer composed of a number of copies of the system (beads), providing a rigorous route to including NQEs in simulations. Despite their accuracy, the use of these approaches is limited owing to their computational cost, which scales linearly with the bead count.

Many strategies have been pursued to accelerate PI simulations.  These include ring-polymer contraction (RPC) \cite{markland_efficient_2008}, modification of the free ring-polymer evolution \cite{,korol2019cayley,korol2020dimension}, advanced integrators (e.g., PIGLET, PILE-G  \cite{ceriotti2010efficient,liu_simple_2016,ceriotti2012efficient, pollock1984pimc}), machine learning force fields \cite{fan2025performing,li2022using}, and software engineering \cite{ceriotti2014_ipi, litman2024pi}. However, as long as the ring polymer remains, the expense exceeds that of standard MD (i.e., non-PIMD). An alternative is to embed NQEs directly into classical dynamics through colored-noise thermostats \cite{ceriotti2009nuclear}, coarse-graining potentials \cite{musil2022quantum,zaporozhets2024accurate} or quantum thermal baths  \cite{dammak2009quantum}. These methods can achieve MD-level costs but break down when systems are strongly anharmonic and/or at low temperatures. In addition, existing approaches generally require rerunning simulations when the ensemble temperature changes.

This cost–accuracy tension motivates a different route: learning a data-driven map that recovers equilibrium quantum statistics at fixed imaginary-time interval $\tau=\beta/P$ (where $\beta=1/k_BT$ is the inverse temperature and $P$ is the number of beads) from standard MD. This strategy has two advantages over those described above. First, once MD data consistent with the target $\tau$ are available, no further force evaluations are needed during sampling. Second, a map trained at a given $\tau$ transfers along a fixed $\tau$-line: changing $T$ while adjusting $P$ to keep $\tau$ constant yields the corresponding finite-$P$ path integral ensemble at temperature $T$, without retraining.

Our approach adopts a Gibbs sampling \cite{gelfand2000gibbs} perspective on the ring-polymer ensemble. The core strategy relies on the single-bead conditional distribution $p_\tau(\mathbf{x}_i\mid\mathbf{x}_{-i})$. Here, $\mathbf{x}_i$ denotes the coordinate of the $i$-th bead, and $\mathbf{x}_{-i}$ represents the configuration of all other beads. Once a high-fidelity sampler for this conditional distribution is available, repeated single-bead Gibbs updates are sufficient to sample the target ring-polymer distribution.
Our key observation is that, since the single-bead conditional distribution depends only on the imaginary-time interval $\tau$ and the midpoint of the two neighboring beads, we can train an accurate generative approximation to the conditional distribution using data from standard MD. % involving only two beads.(Actually we only need one, so I delete this) 
% Our key observation is that the single-bead conditional has a particularly simple structure for a fixed system: it depends strictly on the imaginary-time interval $\tau$ and the midpoint of the two neighboring beads, implying a model trained at a given $\tau$ can be reused at different temperatures by adjusting $P$. 
In practice, a lightweight ($\sim$$10^5$-parameter) generative model is sufficient for high-fidelity sampling of the single-bead conditional distribution. 

Training data for this conditional probability distribution can be obtained in three ways: (i) a Bayesian construction together with unrestrained MD, (ii) restrained MD, and (iii)  PIMD/PIMC.  Options (i) and (ii) both rely only on inexpensive standard MD; option (iii) involves costly simulations, but the data can be %reused to access additional state points that share the same $\tau$ without further simulation.
%That is, the dependence on $\tau$ allows the learned model to be 
transferred across temperatures by adjusting $P$ to keep $\tau$ constant.
We illustrate all three of these possibilities below.

In the following, we first derive the theory of our method, then validate the accuracy and efficiency of GG-PI on Zundel ion, bulk water, and para-hydrogen. While we focus on PI simulations in this paper, we point to several important applications with similar Markov structure that could also benefit from the strategy introduced here.

\section{Methods}
The joint distribution of the ring polymer for distinguishable particles is:
\begin{equation}
    \begin{aligned}
        &\pi_\tau(\mathbf x_1,\ldots,\mathbf x_P)\propto %=\frac{1}{Z_P}
        \exp\!\big[-S_E(\{\mathbf x_k\})\big],
        \\
        &S_E(\{\mathbf{x}_k\}) = \sum_{k=1}^{P} \left[ \frac{1}{2\hbar^2\tau}||\mathbf{x}_k-\mathbf{x}_{k-1}||^2_{\mathbf{M}} + \tau V(\mathbf{x}_k) \right]
    \end{aligned}
    \label{eq:ring-polymer_distribution}
\end{equation}
where $S_E$ is the Euclidean action %and $Z_P$ is the partition function
\cite{chandler1981exploiting,herman_path_1982},
%$\mathbf{x}_k \in \mathbb{R}^{D}$ represents the coordinate of the $k$-th bead in a $D$-dimensional system, and $\tau = \beta/P$ denotes the imaginary-time interval. The 
$||\mathbf{v}||_\mathbf{M}^2 = \mathbf{v}^T\mathbf{M}\mathbf{v}$ is the mass-scaled norm, and $\mathbf{M}$ is the diagonal mass matrix. The ring polymer satisfies the cyclic boundary condition $\mathbf x_{0}\equiv\mathbf x_{P}$.

%TODO: Add a high-level feeling about this algorithm
Eq.~\eqref{eq:ring-polymer_distribution} defines a classical ring-polymer isomorphism of the quantum Boltzmann distribution. Importantly, the Euclidean action is local in the bead index: for a given bead $\mathbf x_i$, the only terms in $S_E$ that depend on $\mathbf x_i$ are the on-site $\tau V(\mathbf x_i)$ and the two harmonic couplings to its neighbors $\mathbf x_{i\pm1}$. Hence, conditioning on all other beads leaves
\begin{equation}
\begin{aligned}
\label{eq:local_conditional_short}
&p_\tau(\mathbf x_i\mid \mathbf x_{-i})
\ \propto\
\exp\!\left[
-\tau V(\mathbf x_i) \right ]
\\
&\cdot\exp \left [ 
-\frac{1}{2\hbar^2\tau}
\Big(\|\mathbf x_i-\mathbf x_{i-1}\|_{\mathbf M}^2+\|\mathbf x_{i+1}-\mathbf x_i\|_{\mathbf M}^2\Big) \right ]    
\end{aligned}
\end{equation}
which implies $p_\tau(\mathbf x_i\mid \mathbf x_{-i})=p_\tau(\mathbf x_i\mid \mathbf x_{i-1},\mathbf x_{i+1})$.
Completing the square gives
\begin{multline}
\label{eq:complete_square_short}
\|\mathbf x_i-\mathbf x_{i-1}\|_{\mathbf M}^2+\|\mathbf x_{i+1}-\mathbf x_i\|_{\mathbf M}^2\\
=2\|\mathbf x_i-\mathbf y_i\|_{\mathbf M}^2+\text{const},
\end{multline}
where
$\mathbf y_i=(\mathbf x_{i-1}  +\mathbf x_{i+1})/2$, and the constant is independent of $\mathbf x_i$. We can thus simplify the single-bead conditional density to the following form:
\begin{align}
    %\centering
    %\begin{aligned}
    p_\tau(\mathbf{x}_i\mid\mathbf{x}_{-i}) &= p_\tau(\mathbf{x}_i\mid\mathbf{y}_i)\\ &\propto \exp{(-\tau V(\mathbf{x}_i))} \mathcal{N}(\mathbf{y}_i,\boldsymbol{\Sigma}_\tau)[\mathbf{x}_i], 
    %\end{aligned}
    \label{eq:single_bead_condition}
\end{align}
where $\mathcal{N}(\mathbf{y}_i,\boldsymbol{\Sigma}_\tau)$ is the probability density function of a normally distributed random variable with mean $\mathbf{y}_i$ and covariance matrix $\boldsymbol{\Sigma}_\tau=(\hbar^2\tau/2)\mathbf{M}^{-1}$.

Eq.~(\ref{eq:single_bead_condition}) shows that the single-bead conditional distribution is a Gaussian distribution centered at $\mathbf y_i$ and reweighted by $e^{-\tau V(\mathbf x_i)}$. The Gaussian factor strongly localizes \(\mathbf x_i\) around \(\mathbf y_i\), so a modest-capacity conditional flow suffices to approximate $p_\tau(\mathbf{x}_i\mid\mathbf{y}_i)$ in practice. This conditional distribution also admits a denoising-posterior interpretation, where the kinetic energy acts as a universal Gaussian noise and the potential energy as the signal. This is also closely related to diffusion models; we briefly discuss this connection in the Appendix \ref{app:denoising}

We target the ring-polymer distribution in Eq.~(\ref{eq:ring-polymer_distribution}) via parallel odd–even Gibbs updates \cite{gelfand2000gibbs} using the single-bead conditional probability distribution in Eq.~(\ref{eq:single_bead_condition}). Repeatedly resampling beads from their exact conditional probability distributions defines a Markov chain whose stationary distribution is $\pi_\tau(\mathbf x_{1:P})$. However, direct inner Markov chain Monte Carlo per bead is slow, so we learn an E(3)-equivariant conditional flow $q_\theta \approx p_\tau$ trained by flow matching \cite{chen2018neural,lipman_flow_2023} and use it to update all odd/even beads in parallel during inference. Metropolis-within-Gibbs \cite{tierney1994markov} can be used to ensure sampling converges to an exact distribution, but we omit it in our experiments due to the empirically observed similarity of the proposal and target distributions.

To obtain training data without path-integral simulation, we apply Bayes' theorem to Eq.~(\ref{eq:single_bead_condition}):
\begin{equation}
    p_\tau(\mathbf{x}_i\mid\mathbf{y}_i)=\frac{p_\tau(\mathbf{x}_i)p_\tau(\mathbf{y}_i\mid\mathbf{x}_i)}{p_\tau(\mathbf{y}_i)},
    \end{equation}
    where we take
    \begin{equation}
        \begin{aligned}
            p_\tau(\mathbf{x}_i)&=\frac{1}{Z_\tau}e^{-\tau V(\mathbf{x}_i)}
            \\
            p_\tau(\mathbf{y}_i\mid\mathbf{x}_i)&=\mathcal{N}(\mathbf{x}_i,\boldsymbol{\Sigma}_\tau)[\mathbf{y}_i].
        \end{aligned}
    \end{equation}

Sampling $p_\tau({\mathbf x}_i)$ is equivalent to running a classical simulation at the inverse temperature $\tau$ or equivalently scaling the potential $V \rightarrow V/P$ at the target temperature. Both broaden coverage and lower barriers. Samples of ${\mathbf y}_i$ can then be sampled easily from $\mathcal{N}(\mathbf{x}_i,\boldsymbol{\Sigma}_\tau)$. Alternatively, ${\mathbf y}_i$ samples can first be drawn from any convenient prior and corresponding samples from $p_\tau(\mathbf{x}_i\mid\mathbf{y}_i)$ can then be generated by restrained MD. 

A key observation is that, although pairs $(\mathbf{x}_i,\mathbf{y}_i)$ constructed in this way are not drawn from the ring--polymer-induced joint marginal of $(\mathbf{x}_i,\mathbf{y}_i)$ implied by Eq.~(\ref{eq:ring-polymer_distribution}), they are still valid training samples for $p_\tau(\mathbf{x}_i\mid\mathbf{y}_i)$ because this construction yields the correct conditional probability distribution.

Finally, when PIMD/PIMC trajectories with a given $\tau$ are already available, one can extract $({\mathbf x}_i,{\mathbf y}_i)$ pairs from them. This provides a convenient way to study other ensembles with the same $\tau$ without rerunning PIMD/PIMC.

Once trained, we initialize multiple independent chains and update them in parallel using the generative approximation of the single-bead conditional distributions. 
% After equilibration, we retain only the terminal state from each chain, yielding effectively independent draws across chains (see SI for details). 

In this paper, we work within the imaginary-time path-integral representation of quantum statistics for distinguishable particles in the canonical ($NVT$) ensemble. Extending GG-PI to indistinguishable particles would require incorporating exchange effects and addressing the sign problem in fermions \cite{hirshberg2019path,hirshberg2020path}, but these possibilities are beyond the scope of the paper. For clarity, we focus on the primitive approximation \eqref{eq:ring-polymer_distribution} in the main text and mention other ring polymer approximations in the SI.

\section{Results and Discussion}

We validate our method on three systems: Zundel ion, bulk water and para-hydrogen followed by a sampling-only performance comparison. In each system, we compare GG-PI against reference PIMD and, as appropriate, standard MD. We exercise all three routes to construct training pairs $(\mathbf{x}_i,\mathbf{y}_i)$ for learning the conditional
$p_\tau(\mathbf{x}_i\mid\mathbf{y}_i)$: PIMD for Zundel ion, restrained MD for water, and the Bayesian construction for para-hydrogen. Full protocols are in the SI.

\subsection{Zundel ion}

\begin{figure}[h]
  \centering
  \includegraphics[width=0.98\linewidth]{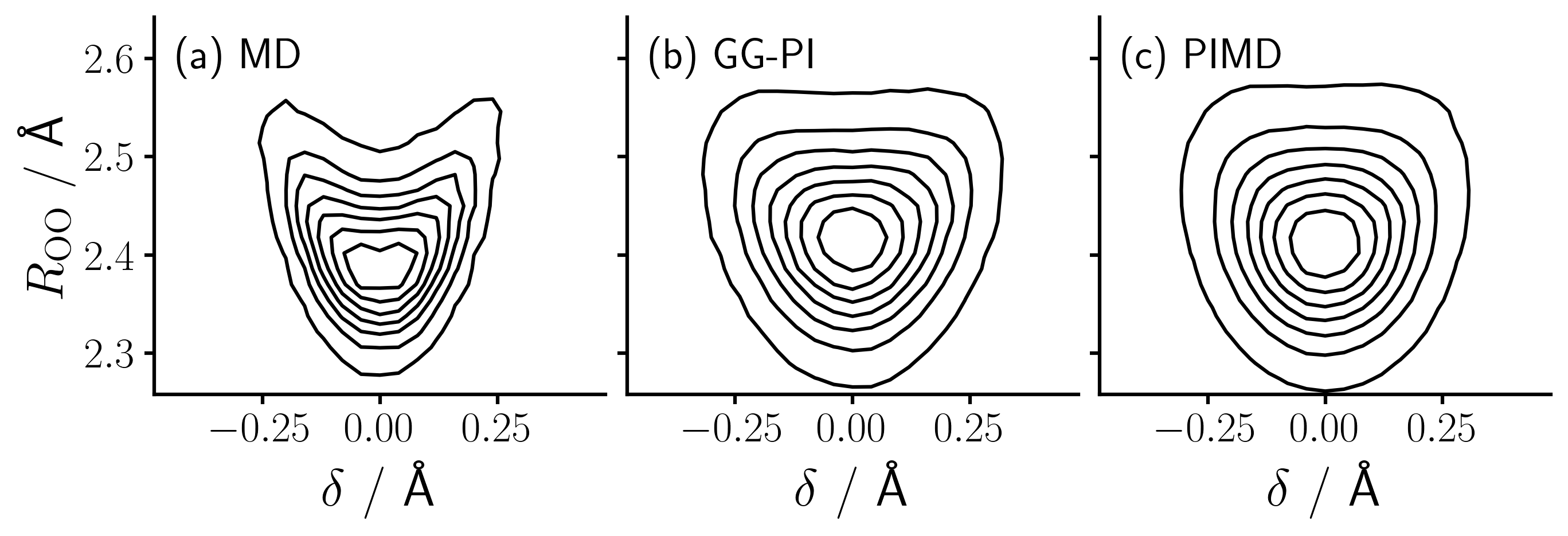}
  \caption{
    Zundel cation (H$_5$O$_2^+$) at $300~\mathrm{K}$. Joint distribution $P(\delta,R_{\mathrm{OO}})$
    with $\delta=R(\mathrm{O_aH})-R(\mathrm{O_bH})$ and $R_{\mathrm{OO}}$.
    Panels: \textbf{(a)} standard MD, \textbf{(b)} GG–PI, and \textbf{(c)} reference PIMD.
    Path–integral simulations use $P=32$; identical contour levels are used across panels.
  }
  \label{fig:zundel_2d_density}
\end{figure}

The Zundel cation, H$_5$O$_2^+$, comprises a shared proton between two water molecules. At $T=300~\mathrm K$, GG-PI reproduces the correct proton-sharing distribution observed by PIMD, whereas standard MD fails  (Fig.~\ref{fig:zundel_2d_density}). Quantitatively, GG-PI captures the significant quantum delocalization of the shared proton, yielding a ring-polymer radius of gyration $R_g \approx 0.17$ \AA. GG-PI also reproduces the quantum structural fluctuations: The standard deviation of the O-O distance is 0.062 \AA\ for GG-PI and PIMD, compared with 0.054 \AA\ for standard MD.

\subsection{Bulk water}

\begin{figure}[t]
  \centering
  \includegraphics[width=0.98\linewidth]{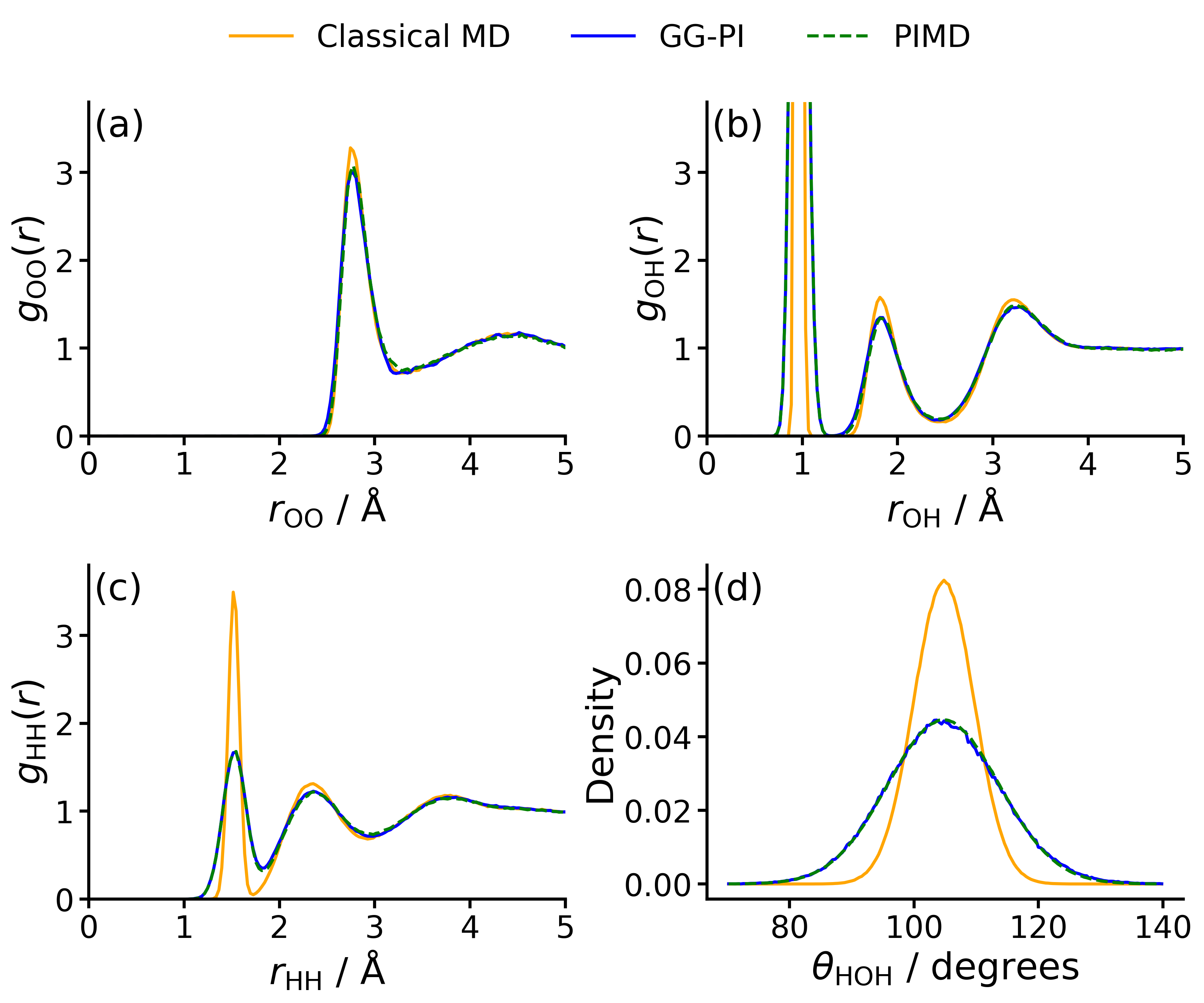}
  \caption{Structural properties of bulk liquid water at 300~K, comparing standard MD (orange solid lines), our GG-PI (blue solid lines), and a reference PIMD simulation (green dashed lines).
        All path-integral simulations are performed with $P=32$.
        The four panels display: \textbf{(a)} the O-O radial distribution function (RDF), \textbf{(b)} the O-H RDF, \textbf{(c)} the H-H RDF, and \textbf{(d)} the intramolecular H-O-H bond angle distribution.}
  \label{fig:water_rdf}
\end{figure}

We next test bulk liquid water at $T=300~\mathrm K$ (216 molecules; Fig.~\ref{fig:water_rdf}). GG-PI matches PIMD for the O–O, O–H, and H–H radial distribution functions (RDFs) and for the intramolecular H–O–H angle, whereas standard MD is over-structured.

% Water figure (single column)

\subsection{Para-Hydrogen}

\begin{figure}[ht]
  \centering
  \includegraphics[width=0.98\linewidth]{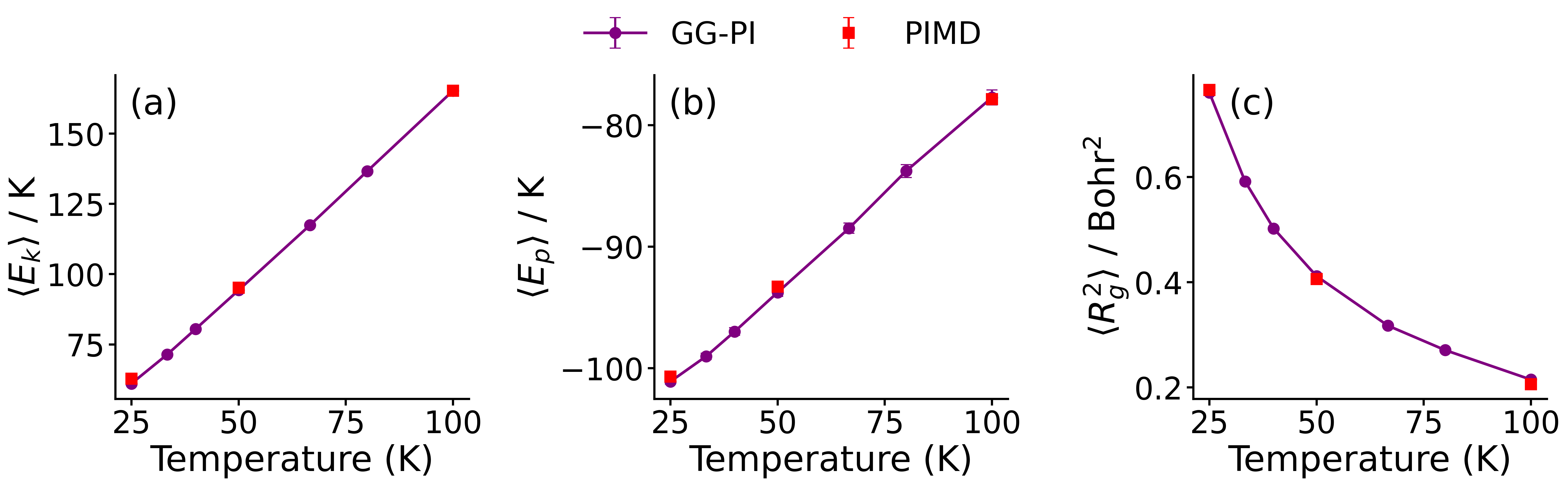}
  \caption{
  Temperature dependence of three observables for para-H$_2$:
  \textbf{(a)} kinetic energy per molecule,
  \textbf{(b)} potential energy per molecule, and
  \textbf{(c)} radius of gyration.
  % GG\mbox{-}PI generates a continuous curve across temperatures after being trained once on a 64-molecule system using restrained MD at $100~\mathrm{K}$; the trained model is then applied, without retraining, to a 172-molecule system at the same density.
  PIMD data are simulated at $25~\mathrm{K}$ ($P=32$), $50~\mathrm{K}$ ($P=16$), and $100~\mathrm{K}$ ($P=8$).
  GG\mbox{-}PI is shown as a purple solid line with circle symbols and PIMD as red square symbols; error bars indicate statistical uncertainties.
  }
  \label{fig:sg_potential_vs_temperature}
\end{figure}

\begin{figure}[ht] 
    \centering
    \includegraphics[width=0.98\linewidth]{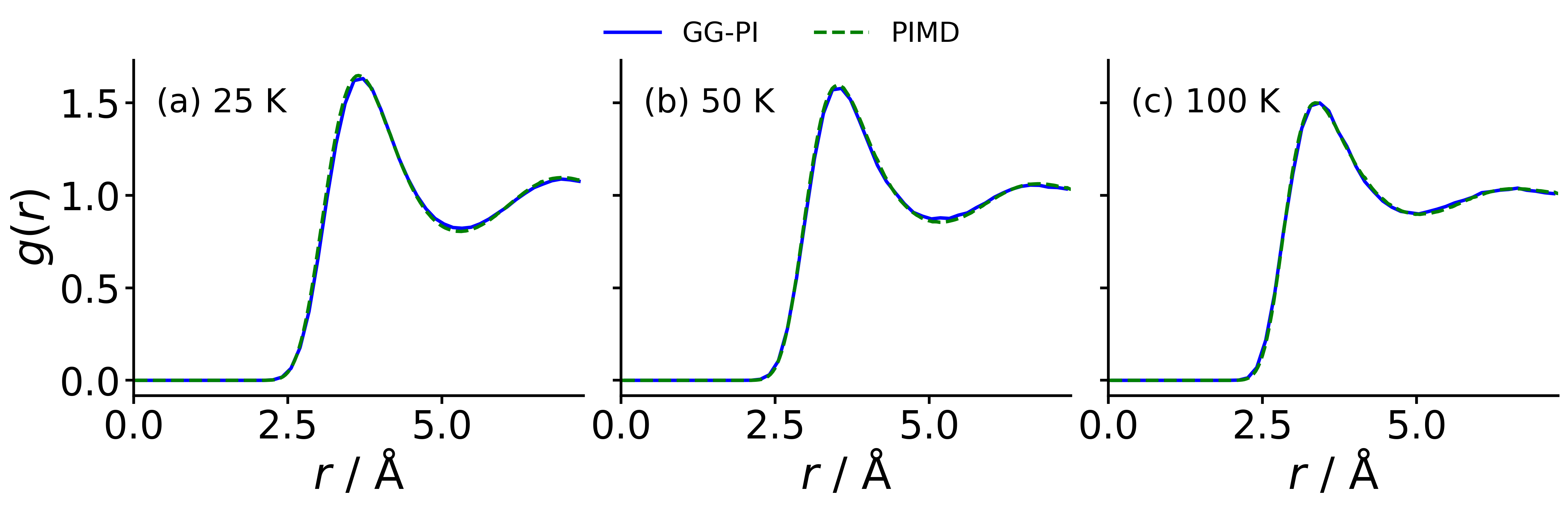}
    \caption{
        Radial distribution functions for para-H$_2$ generated by GG-PI (blue solid lines) against reference PIMD simulation (green dashed lines) at three different temperatures with the same $\tau$: \textbf{(a)} $T$=25~K, $P=32$; \textbf{(b)} $T$=50~K, $P=16$;  \textbf{(c)} $T$=100~K, $P=8$. 
        % The model and PIMD data are the same in the Fig \ref{fig:sg_potential_vs_temperature}. 
    }
    \label{fig:sg_rdf_comparison}
\end{figure}

Finally, we probe transfer along a $\tau$-line in para-hydrogen (para-H$_2$), modeled as distinguishable spherical particles interacting via the Silvera–Goldman potential \cite{silvera_isotropic_1978}. Training pairs are generated via the Bayesian construction by sampling $p_\tau(\mathbf x)\propto e^{-\tau V(\mathbf x)}$ using standard MD at the corresponding effective inverse temperature $\tau$ (here, $T_{\rm eff}=800$~K for $T=100$~K and $P=8$).
 A single conditional generative model trained once on a 64-molecule system is applied, without retraining, to a 172-molecule system at the same density from $(T,P)=(100\ \rm{K},8)$ to $(25 \ \rm{K},32)$ with fixed $\tau=\beta/P$. Energies and radii of gyration track PIMD across $T$ (Fig.~\ref{fig:sg_potential_vs_temperature}), and RDFs agree at all state points (Fig.~\ref{fig:sg_rdf_comparison}). A Rao–Blackwellized estimator (see SI) further reduces variance.

\subsection{Performance}

Table~\ref{tab:performance_comparison} compares sampling efficiency in terms of effective sample size per second. While GG-PI incurs a preparation cost, it is a one-off overhead that is negligible compared to the computational burden of PIMD and can be further reduced if we have existing trajectories. Other performance metrics, such as detailed cost breakdown, are reported in Appendix \ref{app:efficiency} and SI.
For Zundel and liquid water, GG-PI significantly accelerates sampling. This speedup is largely gained from the generative model's ability to bypass force evaluations which is often the bottleneck in PIMD simulation. For para-H$_2$, the smaller speedup reflects two factors: the low computational cost of the analytical Silvera-Goldman potential and the increased number of Gibbs sweeps required when quantum effects are more pronounced. Taken together, these results show that GG-PI is most advantageous for systems with moderate quantum effects and expensive potentials (e.g., \textit{ab initio}). In these cases, the high cost of force evaluation far exceeds the overhead of generative inference.

\begin{table}[htb]
\centering
\setlength{\tabcolsep}{6pt} 
\caption{Performance comparison. ``Overhead'' denotes the preparation time of data collection and training for GG-PI. ``ESS/sec'' denotes the Effective Sample Size per second. Speedup is calculated based on ESS/sec. Hardware: Intel Xeon Gold 6346 CPU (32 cores) and NVIDIA RTX 5090 GPU (32 GB).}
\label{tab:performance_comparison}
% \footnotesize 
\begin{tabular}{l l r r r}
\hline
System & Method & Overhead  & ESS/sec & Speedup \\
\hline
Zundel & PIMD & -- &  3.98 & -- \\
       & GG-PI & $<$ 5 min\textsuperscript{\emph{a}} & \textbf{199.4} & \textbf{50$\times$} \\
\hline
Water  & PIMD & -- &  0.22 & -- \\
       & GG-PI & $\sim$60 min  & \textbf{1.96} & \textbf{8.9$\times$} \\
\hline
para-H$_2$ & PIMD & -- & 2.50 & -- \\
           & GG-PI & $\sim$10 min  & \textbf{3.97} & \textbf{1.6$\times$} \\
\hline
\multicolumn{5}{l}{\scriptsize \textsuperscript{\emph{a}} Training time only.} \\
% \multicolumn{5}{l}{\scriptsize \textsuperscript{\emph{b}} $\tau$ in units of simulation/iteration steps.}
\end{tabular}
\end{table}

\section{Conclusions}

We introduce GG-PI, a data-driven approach that recovers equilibrium quantum statistics at fixed $\tau$ from standard MD. Because the single-bead conditional probability distribution is nearly Gaussian, a compact equivariant flow suffices; training uses only classical data and sampling requires no force evaluations. A single conditional generative model transfers along a fixed-$\tau$ line across temperatures. Across Zundel ion, liquid water, and para-H$_2$, GG-PI matches PIMD structures and energies while reducing wall time. By learning the distribution of a bead with more distant neighbors, we anticipate that sampling time in GG-PI can be further reduced following the multilevel sampling strategies in \cite{weare2007efficient,chorin2008mcnochains,marchand2023genmcnochains}.

While our current scope covers distinguishable particles in the canonical ensemble, the local imaginary-time propagator is independent of topological boundary conditions. This implies that our framework can be seamlessly generalized from closed ring polymers to open chains for Path Integral Ground State (PIGS)\cite{sarsa2000path} simulations. Beyond this, extending the method to address indistinguishable particles \cite{hirshberg2019path} and quantum dynamics \cite{cao1994formulation,craig2004quantum} remains a promising frontier for future research. Furthermore, since classical transition paths share the same Markovian structure, our framework naturally extends to sampling trajectories in broader contexts, such as classical transition path sampling and bridge generation in diffusion processes.

\section*{Acknowledgments}
This work was supported by National Institutes of Health award R35 GM136381 and completed with computational resources administered by the University of Chicago Research Computing Center, including Beagle-3, a shared GPU cluster for biomolecular sciences supported by the NIH under the High-End Instrumentation (HEI) grant program award 1S10OD028655-0. JW's effort was supported by National Science Foundation award 2425899.
%\end{acknowledgments}

\section*{Data and Code Availability}
Code and scripts to reproduce all experiments, along with trained checkpoints and processed datasets, are available at \url{https://github.com/wwz171/GenerativeGibbsSamplingforPathIntegral.git}. Raw trajectories and figure data will be provided upon request or via the repository release.

%\bibliographystyle{apsrev4-2}
%\bibliography{reference}% Produces the bibliography via BibTeX.
%apsrev4-2.bst 2019-01-14 (MD) hand-edited version of apsrev4-1.bst
%Control: key (0)
%Control: author (72) initials jnrlst
%Control: editor formatted (1) identically to author
%Control: production of article title (-1) disabled
%Control: page (0) single
%Control: year (1) truncated
%Control: production of eprint (0) enabled
%

\appendix
\section{Equivalence to Gaussian Denoising}
\label{app:denoising}
Consider the short-time propagator in the coordinate basis: 
\begin{equation}
    \begin{aligned}
    K_\tau(\mathbf{x}_i,\mathbf{x}_{i-1}) &\propto \exp\left [-\frac{\tau}{2}(V(\mathbf{x}_i)+V(\mathbf{x}_{i-1}))\right ] 
    \\
    &\quad \cdot \exp \left ( 
-\frac{1}{2\hbar^2\tau}
\|\mathbf x_i-\mathbf x_{i-1}\|_{\mathbf M}^2 \right ).
    \end{aligned}
\end{equation}
The single-bead conditional distribution is the product of two short-time propagators meeting at $\mathbf{x}_i$ with all other beads fixed:
\begin{equation}
    p_\tau(\mathbf{x}_i|\mathbf{x}_{-i}) \propto K_\tau(\mathbf{x}_{i+1},\mathbf{x}_i)K_\tau(\mathbf{x}_i,\mathbf{x}_{i-1}).
\end{equation}
Since the kinetic part of the propagator is Gaussian, the product of two propagators shares the same structure as a single propagator, differing only in its parameters. Completing the square yields the form derived in the main text:
\begin{equation}
    \begin{aligned}
    p_\tau(\mathbf{x}_i\mid\mathbf{x}_{-i}) &= p_\tau(\mathbf{x}_i\mid\mathbf{y}_i)\\ &\propto \exp{(-\tau V(\mathbf{x}_i))} \mathcal{N}(\mathbf{y}_i,\boldsymbol{\Sigma}_\tau)[\mathbf{x}_i], 
    \label{eq:app_single_bead}
    \end{aligned}
\end{equation}
where $\mathbf{y}_i = (\mathbf{x}_{i-1}+\mathbf{x}_{i+1})/2$ and $\boldsymbol{\Sigma}_\tau=(\hbar^2\tau/2)\mathbf{M}^{-1}$.
Equation~\eqref{eq:app_single_bead} implicitly defines a denoising task. 
Recalling the Bayesian decomposition in Eqs.~(6) and (7) of the main text, the likelihood $p_\tau(\mathbf{y}_i\mid\mathbf{x}_i) = \mathcal{N}(\mathbf{x}_i,\boldsymbol{\Sigma}_\tau)[\mathbf{y}_i]$ represents the physical injection of Gaussian noise arising from the kinetic energy term. 

Consequently, sampling the posterior $p_\tau(\mathbf{x}_i\mid\mathbf{y}_i)$ is mathematically equivalent to a Gaussian denoising problem. To see this connection, consider a diffusion model with a variance-exploding (VE) forward process \cite{ho2020ddpm,song2020score}, where the noise kernel is:
\begin{equation}
    p(\mathbf{x}_t\mid \mathbf{x}_0)
    =
    \mathcal{N}(\mathbf{x}_0,\boldsymbol{\Sigma}_t)[\mathbf{x}_t].
\end{equation}
\begin{table*}[ht]
\centering
\renewcommand{\arraystretch}{1.2}
\setlength{\tabcolsep}{12pt}

\caption{Computational costs and sampling efficiency for GG-PI. ``Data Time'' denotes the wall-clock time to collect the training samples. ``Training Time'' denotes the wall-clock time for model training. ``Burn-in'' refers to the number of steps discarded to reach equilibrium. ``IAT'' denotes the maximum Integrated Autocorrelation Time. Hardware remain the same.}
\label{tab:ggpi_computational_cost}

\begin{tabular}{l c c c c c c}
\hline
System & Training Samples & Data Time & Training Time & Burn-in & IAT\textsuperscript{\emph{b}} & ESS/sec \\

\hline
Zundel & --\textsuperscript{\emph{a}} & -- & $<$ 5 min & 100 & 15 & 199.36 \\
\hline
Water & 10000 & 50 min & 10 min & 50 & 20 & 1.96 \\
\hline
para-H$_2$ & 20000 & 10 min & 5 min & 800 & 40 & 3.97 \\
\hline
\multicolumn{7}{l}{\footnotesize \textsuperscript{\emph{a}} Reusing existing PIMD data; only training time applies.}
\\
\multicolumn{7}{l}{\footnotesize \textsuperscript{\emph{b}} IAT in units of simulation/iteration steps.}
\end{tabular}
\end{table*}

The corresponding denoising posterior is given by:
\begin{equation}
    p(\mathbf{x}_0\mid \mathbf{x}_t)
    \propto
    p(\mathbf{x}_0)\,
    \mathcal{N}(\mathbf{x}_t,\boldsymbol{\Sigma}_t)[\mathbf{x}_0].
\end{equation}
By identifying $\mathbf{y}\equiv \mathbf{x}_t$, $p(\mathbf{x}_0)\equiv p_\tau(\mathbf{x})$, and $\boldsymbol{\Sigma}_t\equiv \boldsymbol{\Sigma}_\tau$, we demonstrate that the single-bead conditional distribution is formally identical to the denoising posterior of a diffusion model at a fixed noise level determined by $\tau$.

\section{Efficiency}
\label{app:efficiency}

To ensure a rigorous and fair comparison of sampling efficiency, we computed the Integrated Autocorrelation Time (IAT) for both GG-PI and PIMD methods. We monitored specific observables to capture different modes of the system dynamics. For the Zundel ion, we tracked the radius of gyration $R_g$, potential energy $U$ and the proton transfer coordinate $\delta$. For water and para-H$_2$, the IAT was measured for $R_g$ and $U$ only. The calculated IAT values are summarized in Table \ref{tab:iat_breakdown}. The maximum IAT among the monitored observables is used to calculate the effective sample size.

\begin{table}[H]
% \centering
% \renewcommand{\arraystretch}{1.0}
\setlength{\tabcolsep}{6pt}
\caption{Integrated Autocorrelation Times (IAT) of different observables. $\delta=R(\mathrm{O_aH})-R(\mathrm{O_bH})$, $U$: potential energy, and $R_g$: radius of gyration. IAT is measured in units of sampling iterations for GG-PI and MD simulation steps for PIMD. The maximum IAT for each system is marked in \textbf{bold}.}
\centering
\label{tab:iat_breakdown}
\begin{tabular}{l l r r}
\hline
System & Observable & IAT\textsubscript{GG-PI} & IAT\textsubscript{PIMD} \\
\hline
Zundel & $\delta$ & \textbf{15} & $<$ 1.0 \\
       & $U$      & 3 & \textbf{30} \\
       & $R_g$    & 7 & 5 \\
\hline
Water  & $U$      & \textbf{20} & \textbf{250} \\
       & $R_g$    & 8 & 7 \\
\hline
para-H$_2$ & $U$  & \textbf{40} & \textbf{100} \\
           & $R_g$ & 35 & 17 \\
\hline
\end{tabular}
\end{table}

Based on the IAT, we measured the Effective Sample Size per second (ESS/s) for each method. Leveraging the parallel execution capabilities of GPUs, we optimized the number of parallel trajectories (batch size) for both architectures to maximize ESS/s. For the CPU, maximum throughput was achieved when each core processed a single independent trajectory. In contrast, the optimal configuration for the GPU varied across different systems (see SI). Table \ref{tab:ggpi_computational_cost} reports the measured ESS/s with a breakdown of training and sampling times.

\clearpage
\onecolumngrid

\setcounter{equation}{0}
\setcounter{figure}{0}
\setcounter{table}{0}
\setcounter{page}{1}
\renewcommand{\theequation}{S\arabic{equation}}
\renewcommand{\thefigure}{S\arabic{figure}}
\renewcommand{\bibnumfmt}[1]{[S#1]}
\renewcommand{\citenumfont}[1]{S#1}

\section*{Supporting Information}

\subsection{Likelihood calculation and Metropolis-Hastings correction}

To generate samples, we define a continuous normalizing flow (CNF) acting on a latent variable $\mathbf{z}_t$. The generative process is governed by the following ODE:
\begin{equation}
\centering
    \begin{aligned}
  \frac{d\mathbf{z}_t}{dt} &= \bm v_\theta(\mathbf{z}_t, \mathbf{y}, t),
  \qquad t\in[0,1];
  \\ 
  \mathbf{z}_0 &\sim \rho_0(\mathbf{z}\mid \mathbf{y}),
  \end{aligned}
  \label{eq:ODE_flow}
\end{equation}
where $\rho_0(\mathbf{z}\mid \mathbf{y})$ is the base distribution, which, in our setting, is a Gaussian distribution centered at $\mathbf{y}$. The generated bead configuration corresponds to the state at $t=1$, i.e., $\mathbf{x} = \mathbf{z}_1$.

The log-density of the generated sample $\mathbf{x}$ (denoted as $\mathbf{z}_1$ in the flow coordinates) satisfies the change of variables formula:
\begin{equation}
\label{eq:cnf_loglik}
  \log q_\theta(\mathbf{z}_1\mid \mathbf{y})
  = \log \rho_0(\mathbf{z}_0\mid \mathbf{y})
  - \int_{0}^{1}\operatorname{Tr}\!\left(\frac{\partial \bm v_\theta}{\partial \mathbf{z}}(\mathbf{z}_t, \mathbf{y}, t)\right)\,dt .
\end{equation}
To guarantee convergence to the exact ring-polymer distribution, a Metropolis-Hastings acceptance criterion can be employed. Denoting the current bead configuration by $\mathbf{x}$ and  the proposal generated by the flow by $\mathbf{x}^\star = \mathbf{z}_1$, the acceptance probability is:
\begin{equation}
\label{eq:mh_alpha}
  \alpha(\mathbf{x}\!\to\!\mathbf{x}^\star\mid \mathbf{y})
  = \min\!\left\{1,\;
    \frac{p_\tau(\mathbf{x}^\star\mid \mathbf{y})}{p_\tau(\mathbf{x}\mid \mathbf{y})}
    \cdot
    \frac{q_\theta(\mathbf{x}\mid \mathbf{y})}{q_\theta(\mathbf{x}^\star\mid \mathbf{y})}
  \right\}.
\end{equation}
Computing $q_\theta(\mathbf{x}\mid \mathbf{y})$ for the current state requires integrating the ODE backward from $\mathbf{z}_1=\mathbf{x}$ to $\mathbf{z}_0$. In our experiments, the CNF proposal matches the target distribution $p_\tau$ closely; thus, we neglect the Metropolis-Hastings step to maximize sampling throughput.

\subsection{Observable and estimator}

In imaginary–time path integrals, for  observables that depend only on position (e.g., potential energy, radial
distribution function, etc.) we write
\begin{equation}
\mathcal O(\mathbf x_{1:P})=\frac{1}{P}\sum_{i=1}^{P}\varphi(\mathbf x_i),
\end{equation}
and the naive estimator over $n$ sampled paths
$\{\mathbf x^{(t)}_{1:P}\}_{t=1}^n$ is
\begin{equation}
\widehat{\mathcal O}_{\text{naive}}
=\frac{1}{n}\sum_{t=1}^{n}\frac{1}{P}\sum_{i=1}^{P}\varphi\!\big(\mathbf x^{(t)}_i\big).
\end{equation}

We reduce variance by \emph{fixing neighbors} $(\mathbf x^{(t)}_{i-1},\mathbf x^{(t)}_{i+1})$ and resampling bead $i$
from its single–bead conditional density $p_\tau(\mathbf x^{(t)}_i\mid \mathbf x^{(t)}_{i-1},\mathbf x^{(t)}_{i+1})$.
With $B$ conditional draws $\{\mathbf x^{(t,m)}_i\}_{m=1}^{B}$ per bead, we form
\begin{equation}
\widehat{m}^{(t)}_i
=\frac{1}{B}\sum_{m=1}^{B}\varphi\!\big(\mathbf x^{(t,m)}_i\big),
\qquad
\widehat{\mathcal O}_{\text{RB}}
=\frac{1}{n}\sum_{t=1}^{n}\frac{1}{P}\sum_{i=1}^{P}\widehat{m}^{(t)}_i .
\end{equation}
By the Rao–Blackwell Theorem,
$\mathrm{Var}[\widehat{\mathcal O}_{\text{RB}}]\le \mathrm{Var}[\widehat{\mathcal O}_{\text{naive}}]$,
while both estimators are unbiased. In practice, a small $B$ already yields a clear variance
reduction.
Specifically, in our calculations we set $B=10$ for the conditional resampling of each bead. 

For the kinetic energy we use the \emph{centroid virial} estimator.
Let $\mathbf x_c=\frac{1}{P}\sum_{i=1}^{P}\mathbf x_i$ denote the centroid.
% Let $\mathbf x_i\!\in\!\mathbb R^{3N}$ denote the $3N$-dimensional configuration at bead $i$
% and $\mathbf x_c=\frac{1}{P}\sum_{i=1}^{P}\mathbf x_i$ the centroid.
With potential $V(\mathbf x_i)$, the total
kinetic energy is \cite{tuckerman_statistical_2023}
\begin{equation}
\label{eq:cv}
\langle K\rangle_{\text{CV}}
= \frac{3N}{2\beta}
+ \frac{1}{2P}\sum_{i=1}^{P}
\big\langle\big(\mathbf x_i-\mathbf x_c\big)\!\cdot\!\nabla V(\mathbf x_i)\big\rangle
\end{equation}

We calculate the ring-polymer radius of gyration, $R_g$, as follows.
Consider an atom $a$ with bead coordinates $\{\mathbf x_{a,k}\}_{k=1}^P$, where $\mathbf x_{a,k} \in \mathbb{R}^3$.
The atomic centroid is defined as
\(
\bar{\mathbf x}_{a}=\frac{1}{P}\sum_{k=1}^{P}\mathbf x_{a,k},
\)
and the per-atom squared radius of gyration at sample step $t$ is given by
\begin{equation}
\label{eq:rg_per_atom}
R_{g,a}^2(t)=\frac{1}{P}\sum_{k=1}^{P}\big\|\mathbf x_{a,k}(t)-\bar{\mathbf x}_{a}(t)\big\|^2 .
\end{equation}
We report the system-level scalar $R_g$ as the mean over all $N$ atoms,
\begin{equation}
\label{eq:rg_system}
R_g(t)=\frac{1}{N}\sum_{a=1}^{N}\sqrt{R_{g,a}^2(t)}.
\end{equation}
% To assess accuracy, we also compute the absolute deviation from a reference value $R_g^{\mathrm{ref}}$, defined as
% \begin{equation}
% \label{eq:rg_diff}
% \Delta R_g(t)=\big|R_g(t)-R_g^{\mathrm{ref}}\big|.
% \end{equation}
Under periodic boundary conditions, we apply the minimal–image convention to all displacements.

\subsection{Parameter settings}

\paragraph{Computational environment.}
All PIMD simulations were performed using the i-PI 3.0 engine~\cite{litman2024pi} with the PILE-G thermostat~\cite{ceriotti2010efficient}. Forces and energies were computed using external drivers: LAMMPS~\cite{thompson2022lammps} for water and para-H$_2$, and the i-PI built-in driver for the Zundel ion. For standard and restrained MD simulations used in comparison and training data generation, the Zundel ion was simulated using i-PI with the bead count set to 1, water was simulated using LAMMPS, and para-H$_2$ was modeled using a verified custom Python code.

\paragraph{Reference PIMD simulations}
We generated reference PIMD simulations for three test systems. The specific system setups are listed in Table~\ref{tab:ref_pimd_details}.
\begin{table}[hbt!]
    \centering
    \caption{Reference PIMD simulation details. The table lists the potential energy surface (PES), system size ($N$ molecules, box length $L$), bead count ($P$), and sampling parameters ($\Delta t$, $\tau_{\mathrm{PILE\text{-}G}}$, $N_{\text{samples}}$ and Total steps).}
    \label{tab:ref_pimd_details}
    \setlength{\tabcolsep}{4pt} 
    \begin{tabular}{llcccccccc}
        \hline
        System & PES & $N$ & $L$ (\AA) & $T$ (K) & $P$ & $\Delta t$ (fs) & $\tau_{\mathrm{PILE\text{-}G}}$ (fs) & $N_{\text{samples}}$ & Total steps \\
        \hline
        Zundel & Built-in \cite{huang2005ab} & 1 & -- & 300 & 32 & 0.50 & 50 & 50,000 & $5\times10^{6}$ \\
        Water & q-TIP4P/F \cite{habershon_competing_2009} & 216 & 18.64 & 300 & 32& 0.25 & 50 & 20,000 & $2\times10^{6}$ \\
        para-H$_2$ & Silvera--Goldman \cite{silvera_isotropic_1978} & 172 & 20.70 & 25/50/100 & 32/16/8 & 1.00 & 25 & 50,000 & $5\times10^{6}$ \\
        \hline
    \end{tabular}
\end{table}

\paragraph{Training data}
Training pairs were derived from different sources. Details are listed in Table~\ref{tab:training_data}.
\begin{table}[hbt!]
    \centering
    \caption{ Training data details. The table lists the source, system size ($N$ molecules, box length $L$), and simulation parameters ($\Delta t$, $N_{\text{samples}}$ and Total steps).  For Water and para-H$_2$, a Langevin thermostat (BAOAB) \cite{leimkuhler2013robust} was used with friction $\gamma = \sqrt{2}/\hbar\tau$.}
    \label{tab:training_data}
    \setlength{\tabcolsep}{12pt}
    \begin{tabular}{llcccccc}
        \hline
        System & Source & $N$ & $L$ (\AA) & $T$ (K) & $\Delta t$ (fs) & $N_{\text{samples}}$ & Total steps \\
        \hline
        Zundel & Reference PIMD & 1 & -- & 300 & -- & 200,000 & -- \\
        Water & Restrained MD\textsuperscript{\emph{a}} & 216 & 18.64 & 9600 & 0.05 & 10,000 & $1\times10^{7}$ \\
        para-H$_2$ & Standard MD & 64 & 14.89\textsuperscript{\emph{b}} & 800 & 0.50 & 20,000 & $2\times10^{7}$ \\
        \hline
        \multicolumn{8}{l}{\footnotesize \textsuperscript{\emph{a}} Initialized from equilibrated 300~K standard MD trajectories.} \\
        \multicolumn{8}{l}{\footnotesize \textsuperscript{\emph{b}} Box size scaled to match the density of the reference PIMD system.} \\
    \end{tabular}
\end{table}

\subsection{Network design and hyperparameter settings}

We parameterize the conditional velocity field of the CNF with an E(3)-equivariant backbone based on ViSNet \cite{wang_enhancing_2024}. The model takes as input the current bead positions $\mathbf x\in\mathbb R^{3N}$, the condition $\mathbf y=(\mathbf x_{i-1}+\mathbf x_{i+1})/2$, the scalar time $t\in[0,1]$, and atomic numbers $Z$. By construction,
\[
v_\theta(g\!\cdot\!\mathbf x,\,g\!\cdot\!\mathbf y,\,t)=g\!\cdot\!v_\theta(\mathbf x,\mathbf y,t)\quad
\text{for any }g\in \mathrm{E}(3),
\]
the output velocity is equivariant.

\paragraph{Velocity network}
Our goal is to predict the per-atom velocity
\begin{equation}
\mathbf v \;=\; v_\theta(\mathbf x,\mathbf y,t)\in\mathbb R^{3N},
\end{equation}
using ViSNet’s invariant/vector streams. The data flow in each forward pass is:

\begin{enumerate}
  \item \textbf{Graph construction.} Build a neighbor graph on nodes (atoms) with positions $\mathbf x$ using a distance cutoff $r_c$ (minimum image convention). 
  \item \textbf{Vector seed (equivariant).} Initialize the vector channel with the relative displacement
  \[
  \tilde{\mathbf c}^{(0)}_i = W_{\mathrm{vec}}(\mathbf{x}_i-\mathbf{y}_i)
  \]
  where $W_{\mathrm{vec}}$ is a learned linear map to the chosen vector-channel width.
  \item \textbf{Invariant conditioning.} 
  (i) Embed atomic numbers: $e_Z = \mathrm{Embed}(Z)$; 
  (ii) Time embedding: The scalar time $t$ is mapped to a feature vector $e_t$ using sinusoidal embeddings; 
  (iii) use a linear layer to get the initial invariant feature $\mathbf{h}_i = W_{\text{inv}}[e_Z,e_t]$.
  \item \textbf{Backbone updates.} Stack $L$ ViSNet blocks. Each block mixes invariant and vector channels via equivariant message passing/attention.
  \item \textbf{Equivariant readout.} Apply a linear head to the final vector channels to produce per-atom velocities,
  \[
  \mathbf v_i \;=\; W_{\mathrm{out}}\;\tilde{\mathbf c}^{(L)}_i,
  \]
\end{enumerate}

\paragraph{Training objective and solver.}
We train with \emph{flow matching} on the linear probability path $\mathbf x_t=(1-t)\mathbf x_0+t\mathbf x_1$, minimizing
\[
\mathcal L_{\mathrm{FM}}=\mathbb E\big\|\mathbf v_\theta(\mathbf x_t,\mathbf y,t)-(\mathbf x_1-\mathbf x_0)\big\|^2,
\]
where $\mathbf x_0\!\sim\!\mathcal N(\mathbf y,\Sigma_\tau)$ and $\mathbf x_1$ is a target conditional sample. During inference we integrate $\dot{\mathbf x}_t=\mathbf v_\theta(\mathbf x_t,\mathbf y,t)$ with an explicit Heun integrator using $S=3$ uniform substeps; edge lists are kept fixed over the ODE to reduce overhead.  In the test, we find the number of integrator steps $S$ can range from $3$ to $10$ while keeping a highly accurate results. However, more steps makes refinement fluctuate less while fewer makes sampling faster.

\paragraph{Hyperparameters.}
Table~\ref{tab:hparams_model} lists the backbone configuration per system. Cutoffs are given in Bohr radii to match our input units.

\begin{table}[h]
    \centering
    \caption{Backbone hyperparameters for the velocity network (ViSNet).}
    \label{tab:hparams_model}
    \begin{tabular}{c|c|c|c|c|c|c|c}
        \hline
        System & Layers $L$ & Fourier dim &Hidden dim & Heads & Cutoff (Bohr) & Optimizer\& LR &Batch/Epoch\\
        \hline
        Zundel & 2 & 24& 24 & 4 & 10 & Adam(1e-3) & 1024/10 \\
        water & 3 & 32 & 32& 4 & 6 & Adam(1e-3) & 32/10\\
        para-H$_2$ & 2 & 24& 24 & 4 & 8 & Adam(1e-3) & 256/10\\
        \hline
    \end{tabular}
\end{table}

\subsection{Different factorizations}
In the main text, we only include the primitive approximation. Here, we extend our method to any factorization. Starting with a generic \textbf{Markovian} factorization:
\begin{equation}
\label{eq:pi-mcolor}
\pi_\tau(\mathbf x_{1:P})
\ \propto\ \prod_{k=1}^{P} K^{(c_k)}_\tau(\mathbf x_k,\mathbf x_{k-1}),
\end{equation}
where $c_k\in\{1,\dots,m\}$ is a periodic color assignment with period $m$. The kernel of each color has the following form:
\begin{equation}
\label{eq:kernel-generic}
K^{(c)}_\tau(\mathbf x,\mathbf y)
\propto
\exp\!\Big[-\tfrac{1}{2}\,(\mathbf x-\mathbf y)^\top \mathbf Q^{(c)}_\tau\,(\mathbf x-\mathbf y)\Big]\,
\exp\!\Big[-\tau\,\phi^{(c)}_\tau(\mathbf x,\mathbf y)\Big],
\end{equation}
Notably, common factorizations share the same kinetic propagator $\mathbf{Q}_\tau^{(c)} = \mathbf{M}/\hbar^2\tau$. Substituting this explicit form into Eq.~\eqref{eq:kernel-generic}, this single-bead conditional distribution for the $i$-th bead is:
\begin{equation}
    \begin{aligned}
            p_\tau(\mathbf x_i \mid \mathbf x_{-i}) &\propto\
K^{(c_i)}_\tau(\mathbf x_i,\mathbf x_{i-1})\;
K^{(c_{i+1})}_\tau(\mathbf x_{i+1},\mathbf x_i) 
\\
    &\propto \exp\left[-\frac{1}{\hbar^2\tau}\left\|\mathbf{x}_i-\frac{\mathbf{x}_{i-1}+\mathbf{x}_{i+1}}{2}\right\|_\mathbf{M}^2\right]
    \cdot \exp[-\tau(\phi_\tau^{(c_i)}(\mathbf{x}_{i},\mathbf{x}_{i-1})+\phi_\tau^{(c_{i+1})}(\mathbf{x}_{i+1},\mathbf{x}_{i}))]
    \end{aligned}
    \label{eq:conditional_kernel_single}
\end{equation}

The single-bead condition in Eq.\
 \eqref{eq:conditional_kernel_single} only depends on its two neighboring beads, and it also depends only on $\tau$.

Furthermore, if the function $\phi(\mathbf{x},\mathbf{y})$ is separable:

\begin{equation}
    \phi^{(c)}_\tau(\mathbf{x},\mathbf{y})=\frac{1}{2}(\psi_\tau^{(c)}(\mathbf{x})+\psi_\tau^{(c)}(\mathbf{y}))
\end{equation}
the single-bead condition becomes:
\begin{equation}
    \begin{aligned}
            p_\tau(\mathbf x_i \mid \mathbf x_{-i})
    &\propto \exp[-\frac{1}{2}(\mathbf{x}_i-\mathbf{y}_i)^\top\mathbf{\Sigma}_\tau^{-1}(\mathbf{x}_i-\mathbf{y}_i)] 
    \cdot \exp[-\tau\Psi_{\text{eff}}(\mathbf{x}_i)]
        \end{aligned}
    \label{eq:conditional_kernel_simple}
\end{equation}
where $\mathbf{y}_i = (\mathbf{x}_{i-1}+\mathbf{x}_{i+1})/2$,  $\mathbf{\Sigma}_\tau=\frac{\hbar^2\tau}{2}\mathbf{M}^{-1}$ and $\Psi_{\text{eff}}(\mathbf{x}_i)=\frac{1}{2}(\psi_\tau^{(c_i)}(\mathbf{x_i})+\psi_\tau^{(c_{i+1})}(\mathbf{x_i}))$.

Some common factorizations are listed below:
\begin{itemize}
\item \textit{Primitive.} $m{=}1$ with
\[
\phi^{(1)}_\tau(\mathbf x,\mathbf y)=\tfrac{1}{2}\big[\psi^{(1)}(\mathbf x)+\psi^{(1)}(\mathbf y)\big],
\qquad \]
\[
\psi^{(1)}(\mathbf x)=V(\mathbf x).
\]
Thus \eqref{eq:conditional_kernel_simple} holds with $c_i{=}1$.

\item \textit{Takahashi-Imada.} $m{=}1$. Still endpoint-separable:
\[
\phi^{(1)}_\tau(\mathbf x,\mathbf y)=\tfrac{1}{2}\big[\psi^{(1)}_\tau(\mathbf x)+\psi^{(1)}_\tau(\mathbf y)\big],
\quad
\]
\[
\psi^{(1)}_\tau(\mathbf x)=V(\mathbf x)\;+\;\kappa_{\rm TI}\,\tau^2\hbar^2\,\|\nabla V(\mathbf x)\|^2_{\mathbf M^{-1}}.
\]

\item \textit{Suzuki-Chin.} $m{=}2$ with odd/even colors; endpoint-separable in both colors:
\[
\phi^{(c)}_\tau(\mathbf x,\mathbf y)=\tfrac{1}{2}\big[\psi^{(c)}_\tau(\mathbf x)+\psi^{(c)}_\tau(\mathbf y)\big],
\]
with
\[
\psi^{(\mathrm{even})}_\tau(\mathbf x)=w_{\rm e}\,V(\mathbf x),
\qquad
\]
and
\[
\psi^{(\mathrm{odd})}_\tau(\mathbf x)=w_{\rm o}\!\left[V(\mathbf x)+\kappa_{\rm SC}\,\tau^2\hbar^2\,\|\nabla V(\mathbf x)\|^2_{\mathbf M^{-1}}\right].
\]
Thus \eqref{eq:conditional_kernel_simple} holds with $c_i\in\{\mathrm{even},\mathrm{odd}\}$ selecting the corresponding on-site term.
\end{itemize}

To be compatible with factorization with multiple colors, an $m$-color Gibbs sampling is naturally an alternative to the current odd-even Gibbs sampling where beads of the same color are updated in parallel. The algorithm is given below:
\begin{algorithm}
    \caption{$m$-color Block Gibbs sampling for Imaginary-Time Path Integral}
    \label{alg:mcolor-gibbs}
    \KwIn{Initial path configuration $\{\mathbf x_k^{(0)}\}_{k=1}^P$; color pattern $\{c_k\}$ with period $m$}
    \KwOut{A sequence of path configurations sampled from $\pi_\tau(\mathbf x_{1:P})$}

    \tcp{Cyclic boundary (ring polymer): indices are taken modulo $P$.}
    \tcp{$\mathbf x_{0}^{(t)} \equiv \mathbf x_{P}^{(t)}$, \quad $\mathbf x_{P+1}^{(t)} \equiv \mathbf x_{1}^{(t)}$ for any iteration $t$.}

    \tcp{Color schedule: in sweep $t$, visit colors $c=1,\dots,m$. Sites of the same color are conditionally independent and can be updated in parallel.}
    \tcp{Neighbor rule: when updating site $i$ of color $c$, use the \emph{latest available} neighbors
    $\mathbf x_{i-1}^{\mathrm{lat}}=\mathbf x_{i-1}^{(t)}$ if $c_{i-1}<c$ else $\mathbf x_{i-1}^{(t-1)}$,
    and
    $\mathbf x_{i+1}^{\mathrm{lat}}=\mathbf x_{i+1}^{(t)}$ if $c_{i+1}<c$ else $\mathbf x_{i+1}^{(t-1)}$.}

    \For{$t=1, \dots, N_{\text{steps}}$}{
        \For{$c=1, \dots, m$}{
            \tcp{Update all color-$c$ beads in parallel}
            \For{$i \in \{1,\dots,P\ \mid\ c_i=c\}$}{
                Draw $\mathbf x_i^{(t)} \sim
                p_\tau\!\left(\mathbf x_i \,\middle|\, \mathbf x_{i-1}^{\mathrm{lat}},\, \mathbf x_{i+1}^{\mathrm{lat}}\right)$\;
                \tcp{Use Eq.~\eqref{eq:conditional_kernel_single} or, if separable, Eq.~\eqref{eq:conditional_kernel_simple}.}
            }
        }
    }
\end{algorithm}

\subsection{Throughput}
\begin{figure}[p]
    \centering
    \includegraphics[width=0.95\linewidth]{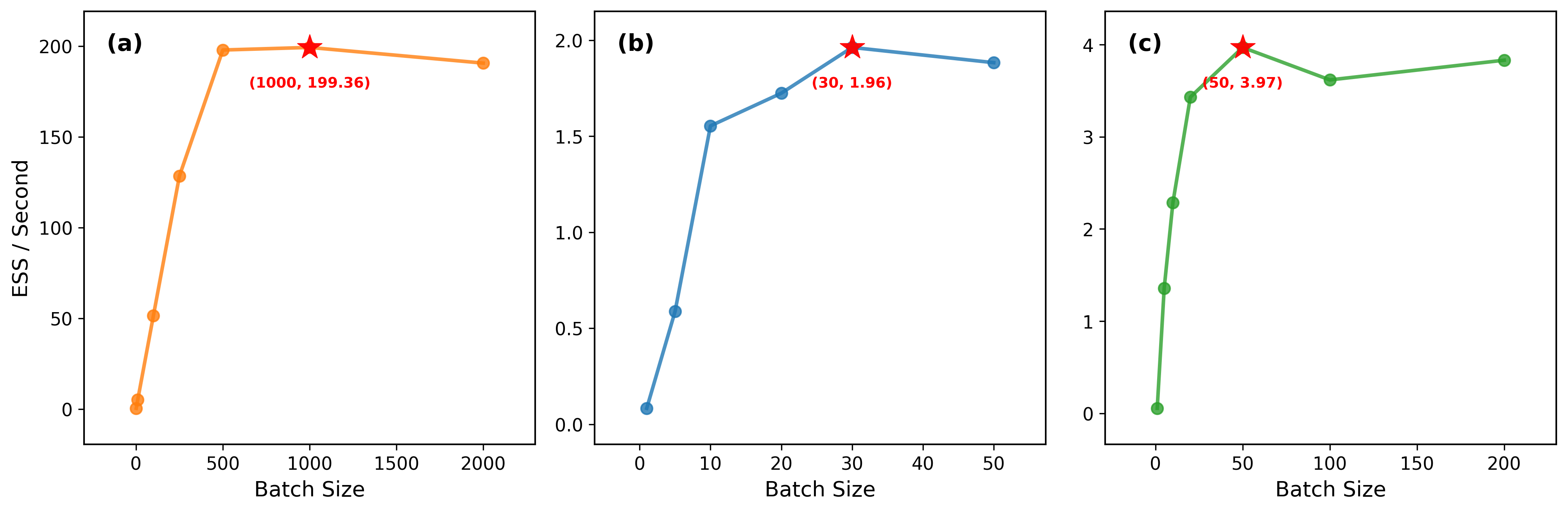} 
    \caption{\textbf{Batch size vs. ESS/sec for GG-PI.} The scanning results for (a) the Zundel cation, (b) water, and (c) para-H$_2$. The red stars indicate the optimal batch sizes that yield the highest sampling efficiency on an NVIDIA RTX 5090 GPU.}
    \label{fig:ess_ggpi}
\end{figure}

\begin{figure}[p]
    \centering
    \includegraphics[width=0.95\linewidth]{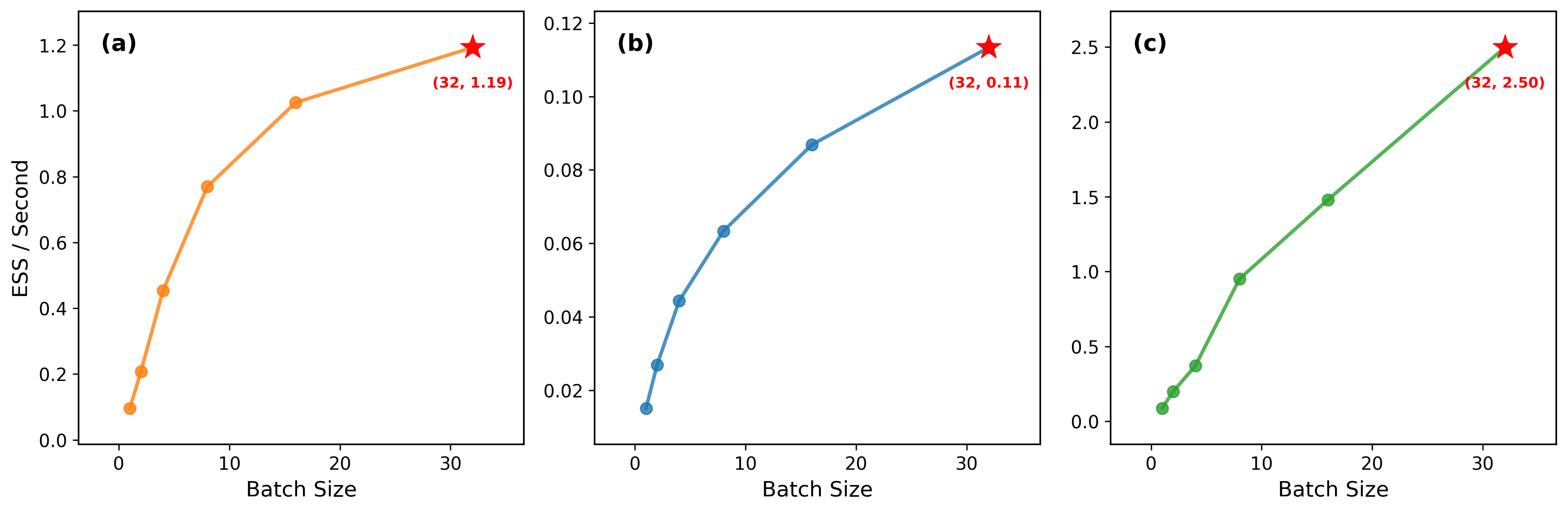} 
    \caption{\textbf{Batch size vs. ESS/sec for PIMD.} The scanning results for (a) the Zundel cation, (b) water, and (c) para-H$_2$. For PIMD, the batch size corresponds to the number of parallel trajectories. Maximal throughput (ESS/sec) is achieved when the batch size matches the number of cores, effectively assigning one trajectory to each core. (Intel Xeon Gold 6346).}
    \label{fig:ess_pimd}
\end{figure}

Here, we measure the throughput of PIMD/GG-PI on CPU/GPU respectively. For PIMD, the batch size denotes the number of parallel trajectories for given CPU, and the maximum throughput is achieved when each core processes a single-independent trajectory. For GPU, this value varies across different systems (Figs. \ref{fig:ess_ggpi} and \ref{fig:ess_pimd}).

\begin{figure}[p]
    \centering
    \includegraphics[width=0.98\linewidth]{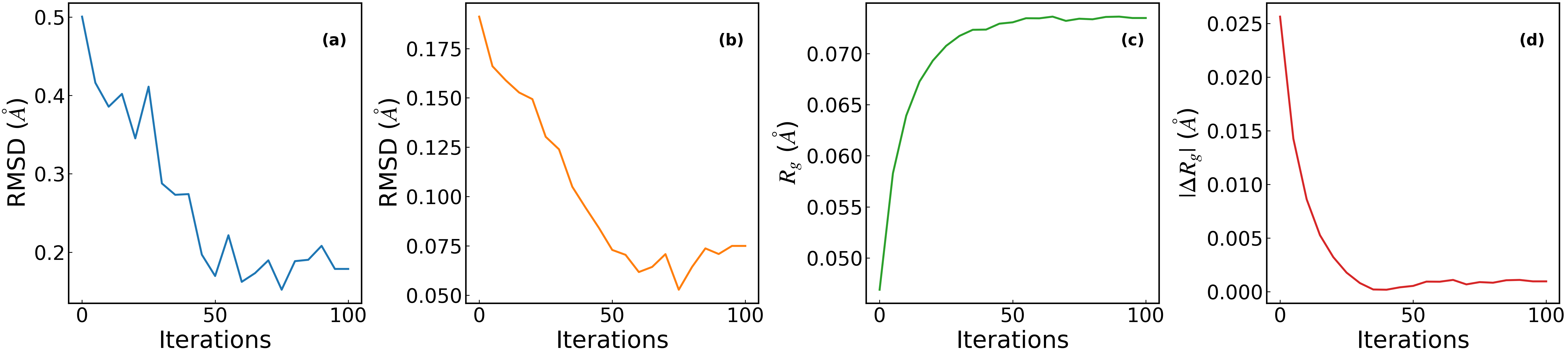}
    \caption{\textbf{Convergence statistics for the Zundel ion system.} 
    (a) Root mean square deviation (RMSD) of the O-O radial distribution function (RDF). 
    (b) RMSD of the Proton RDF. 
    (c) Evolution of the radius of gyration ($R_g$). 
    (d) Absolute difference of the radius of gyration ($|\Delta R_g|$) with respect to the reference. 
    All values are in \aa ngstroms ($\text{\AA}$).}
    \label{fig:ct_zundel}
\end{figure}

\begin{figure}[p]
    \centering
    \includegraphics[width=0.98\linewidth]{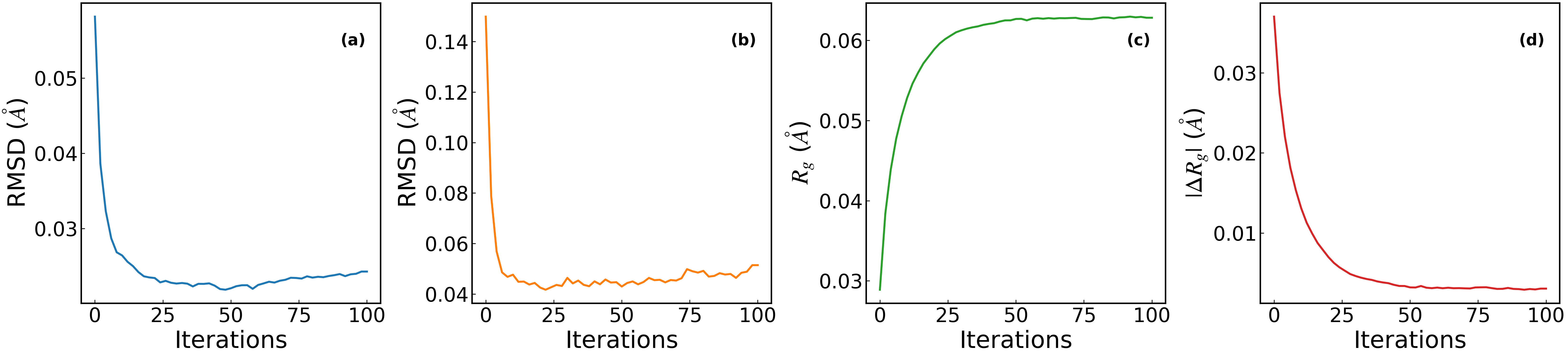}
    \caption{\textbf{Convergence statistics for the bulk water system.} 
    (a) RMSD of the H-H RDF. 
    (b) RMSD of the O-H RDF. 
    (c) Evolution of the radius of gyration ($R_g$). 
    (d) Absolute difference of the radius of gyration ($|\Delta R_g|$) with respect to the reference. 
    All values are in Angstroms ($\text{\AA}$).}
    \label{fig:ct_water}
\end{figure}

\begin{figure}[p]
    \centering
    \includegraphics[width=0.98\linewidth]{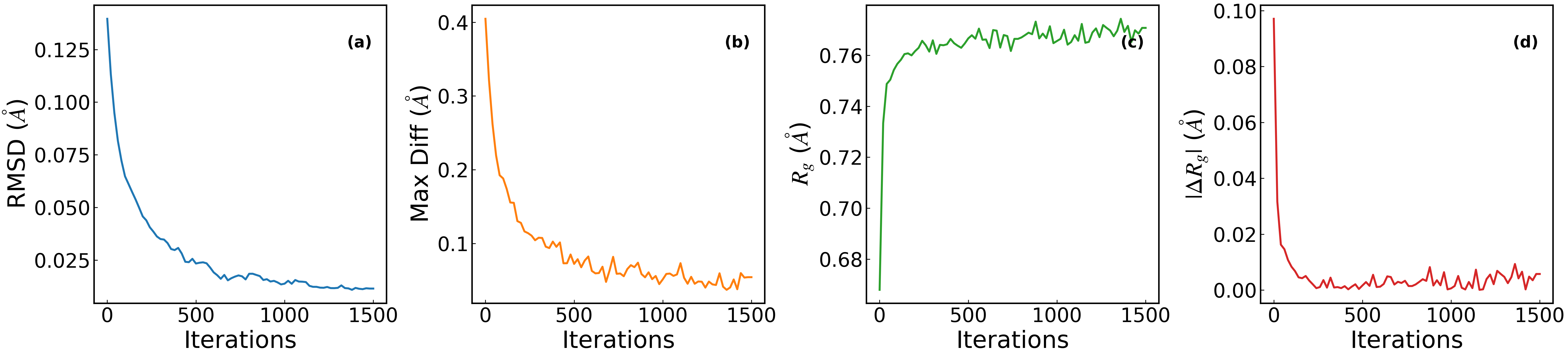}
    \caption{\textbf{Convergence statistics for the Para-hydrogen  system.} 
    (a) RMSD of the radial distribution function (RDF). 
    (b) Maximum pointwise difference (Max Diff) of the RDF. 
    (c) Evolution of the radius of gyration ($R_g$). 
    (d) Absolute difference of the radius of gyration ($|\Delta R_g|$) with respect to the reference. 
    All values are in Angstroms ($\text{\AA}$).}
    \label{fig:ct_sg}
\end{figure}

To determine the burn-in period of GG-PI for each system, we monitored the convergence of structural observables. We tracked the RMSD and maximum absolute deviation between the current radial distribution function (RDF) and the reference PIMD RDF. Additionally, we monitored the ring-polymer radius of gyration $R_g(t)$ and its absolute deviation from the reference, calculated as $\Delta R_g(t)=\big|R_g(t)-R_g^{\mathrm{ref}}\big|$. The convergence curves are presented in Figs.~\ref{fig:ct_zundel}, \ref{fig:ct_water}, and \ref{fig:ct_sg} for the Zundel, water, and para-H$_2$ systems.

\clearpage

\end{document}